\documentclass[12pt,letterpaper]{article}
\usepackage[utf8]{inputenc}

\usepackage{color}
\definecolor{darkblue}{rgb}{0.1,0.1,.7}
\usepackage[colorlinks, linkcolor=darkblue, citecolor=darkblue, urlcolor=darkblue, linktocpage]{hyperref}
\usepackage[]{amsmath}
\usepackage[]{graphicx}
\usepackage[]{latexsym}
\usepackage[utf8]{inputenc}
\usepackage{slashed,graphicx,color,amsmath,amssymb}
\usepackage[mathscr]{eucal}
\usepackage{mathrsfs}
\usepackage[margin=10pt,font=small,labelfont=bf]{caption}
\usepackage{amscd}
\usepackage{bm}
\usepackage{xcolor}
\usepackage{upgreek}
\usepackage[square, comma, sort&compress,numbers]{natbib}
\usepackage[all,cmtip]{xy}
\usepackage[margin=1.7in]{geometry}
\usepackage{cleveref}
\usepackage[color=cyan!30!white,linecolor=red,textsize=footnotesize]{todonotes}
\usepackage{soul}
\numberwithin{equation}{section}
\hypersetup{
	pdfstartview={FitH},    %
	pdftitle={Geometrizing TTbar},    %
	pdfauthor={Caputa, Datta, Jiang, Kraus},     %
	colorlinks=true,       %
	linkcolor=blue,          %
	citecolor=blue,%
	filecolor=magenta,      %
	urlcolor=darkblue           %
}

\newcommand{\tr}{\mathrm{Tr}\,}

\newcommand{\rd}{\mathrm{d}}

\newcommand{\rt}{\mathrm{t}}

\newcommand{\zb}{\bar{z}}
\newcommand{\eps}{\epsilon}
\newcommand{\p}{\partial}

\def\ttb{T{\bar{T}}}

\def\pd{\partial}

\def\bz{{\bar{z}}}

\def\ie{{\it i.e.~}}
\def\eg{{\it e.g.~}}

\def\nn{\nonumber}
\def\pd{\partial}

\def\l1{{{1-loop}}}

\def\bz{{\bar{z}}}

\def\n1{\Bigg|_{n=1}}

\def\n{{(n)}}
\def\tr{{Tr}}

\def\TT{\tilde{\mathcal{T}}}

\def\tr{\text{Tr}}

\def\pd{\partial}

\def\bz{{\bar{z}}}

\def\beq{\begin{equation}}
\def\eeq{\end{equation}}

\def\bea{\begin{eqnarray}}
\def\eea{\end{eqnarray}}
\def\nn{\nonumber}
\def\pd{\partial}

\def\l1{{\text{1-loop}}}

\def\bz{{\bar{z}}}

\def\n1{\Bigg|_{n=1}}

\def\n{{(n)}}
\def\tr{\text{Tr}}

\def\TT{\tilde{\mathcal{T}}}

\def\bz{\bar{z}}

\def\be{\begin{equation}}
\def\ee{\end{equation}}
\def\bal{\begin{array}{l}}
\def\ba#1{\begin{array}{#1}}  %
	\def\ea{\end{array}}
\def\bea{\begin{eqnarray}}
\def\eea{\end{eqnarray}}
\def\beas{\begin{eqnarray*}}
	\def\eeas{\end{eqnarray*}}

\def\nn{\\\nonumber}

\def\ttb{{T\bar{T}}}

\def\eps{\epsilon}

\def\nn{\nonumber}
\def\bit{\begin{item}}
	\def\eit{\end{item}}
\def\benu{\begin{enumerate}}
	\def\eenu{\end{enumerate}}
\def\tr{{\rm tr}}

\def\bz{\bar{z}}

\def\TT{$T\bar{T}$}
\def\OTT{\mathcal{O}_{T\bar{T}}}
\def\lam{\hat{\lambda}}
\def\rt{\rightarrow}

 \makeatletter
 \g@addto@macro\bfseries{\boldmath}
 \makeatother
\makeatletter
\if@todonotes@disabled

\else

\fi
\makeatother

\begin{document}

\definecolor{tinge}{RGB}{255, 244, 195}
\sethlcolor{tinge}
\setstcolor{red}

\vspace*{-.5in} \thispagestyle{empty}
\begin{flushright}
\texttt{CERN-TH-2020-188}
\end{flushright}
\vspace{.2in} {\Large
\begin{center}
{\LARGE \bf
Geometrizing $T\bar{T}$}
\end{center}}
\vspace{.2in}
\begin{center}
{Pawel Caputa$^1$, Shouvik Datta$^{2,3,4}$, Yunfeng Jiang$^{4,5,6}$ and Per Kraus$^2$}
\\
\vspace{.3in}
\small{
$^1$  \textit{Faculty of Physics,
	University of Warsaw, ul.\,Pasteura 5, 02-093 Warsaw, Poland.}
\\ \vspace{.1cm}
\vspace{.2cm}
$^2$ \textit{Mani L.~Bhaumik Institute for Theoretical Physics, Department of Physics \& Astronomy,\\
	 University of California,
	Los Angeles, CA 90095, USA.}
}\\ \vspace{.1cm}
\vspace{.2cm}
$^3$ \textit{Institut f\"ur Theoretische Physik, ETH Z\"urich,\\
	Wolfgang Pauli Strasse 27, CH-8093 Z\"urich, Switzerland.}
\\ \vspace{.2cm}
$^4$ \textit{Department of Theoretical Physics, CERN,\\
	1 Esplanade des Particules, Geneva 23, CH-1211, Switzerland.}
\\ \vspace{.2cm}
$^5$ \textit{Shing-Tung Yau Center of Southeast University, Nanjing 210096, China.}\\  \vspace{.2cm}
$^6$ \textit{School of Physics, Southeast University, Nanjing 211189, China.}

\end{center}

\vspace{.5in}

\begin{abstract}
\normalsize{The $T\bar{T}$ deformation can be formulated as a dynamical change of coordinates. We establish and generalize this relation to curved spaces by coupling the undeformed theory to 2d gravity. For curved space the dynamical change of coordinates is supplemented by a dynamical Weyl transformation. We also sharpen the holographic correspondence to cutoff AdS$_3$ in multiple ways. First, we show that the action of the annular region between the cutoff surface and the boundary of AdS$_3$ is given precisely by the $T\bar{T}$ operator integrated over either the cutoff surface or the asymptotic boundary. Then we derive dynamical coordinate and Weyl transformations directly from the bulk. Finally, we reproduce the flow equation for the deformed stress tensor from the cutoff geometry. }
\end{abstract}

\vskip 1cm \hspace{0.7cm}

\newpage

\setcounter{page}{1}

\noindent\rule{\textwidth}{.1pt}\vspace{-1.2cm}
\begingroup
\hypersetup{linkcolor=black}
\tableofcontents
\endgroup
\noindent\rule{\textwidth}{.2pt}

\section{Introduction}

\TT\ deformed quantum field theories \cite{Smirnov:2016lqw,Cavaglia:2016oda} have many remarkable properties (see \emph{e.g.} \cite{Jiang:2019hxb,Giveon:2019fgr} and references therein) --- particularly in comparison to generic non-renormalizable theories --- yet their status as well-defined quantum theories remains  murky.   It is not  clear whether the standard Wilsonian picture of a quantum field theory as a flow between UV and IR fixed points applies in the case of \TT, since the short distance structure of these theories is not well understood.   An optimistic read on the situation is that the solvable nature of certain observables  in \TT\ deformed theories indicates that there is a well-defined underlying structure, which might turn out to enlarge our understanding of the space of quantum field theories.

It has proven to be useful and illuminating  to view a \TT\ deformed theory as living on a different spacetime metric than its undeformed counterpart \cite{Dubovsky:2017cnj,Cardy:2018sdv,Conti:2018tca,Tolley:2019nmm}.  The simplest context to appreciate this is at the level of classical field theory, where to define a \TT\ deformed theory one is instructed to write down a 1-parameter family of  actions obeying $\frac{dS_\lambda}{d\lambda } = \int d^2x \, \OTT$, where the operator $\OTT\sim \det T^\mu_\nu$ is built out of the stress tensor of the deformed theory.  As we review below, the solution to this problem can be obtained by introducing Hubbard-Stratonovich fields, where these auxiliary fields play the role of the metric which couples to the undeformed theory.   One then solves the equations of motion for these fields and plugs back in to arrive at the deformed Lagrangian.

The two metrics just discussed are related by a coordinate transformation --- in general supplemented by a Weyl transformation --- that depends explicitly on the dynamical fields of the theory.    This notion of a field dependent coordinate (and Weyl) transformation may turn out to hold a key to a deeper understanding of \TT\ deformation, and so our goal in this paper is to develop this idea further.

At the quantum level, the appearance of such  dynamical coordinate transformations can be understood from several points of view.  One route is via the computation of correlation functions in the deformed theory.  The approach taken by Cardy \cite{Cardy:2018sdv,Cardy:2019qao} uses that $\OTT$ can be written as a total derivative away from the insertion of other operators.   Using this, a change in the \TT\ coupling can be absorbed into a change in the operators appearing in the correlator, and this change turns out to correspond to a coordinate transformation that depends explicitly on the stress tensor.    This coordinate transformation agrees with the one discussed above at the level of classical actions.   It should be noted that this construction is so far restricted to \TT\ deformed theories defined on flat space; even defining the $\OTT$ operator on curved space is a challenge, as we review in the main text.

As first discussed by \cite{McGough:2016lol}, it is interesting to view \TT\ deformed theories in the context of the AdS/CFT correspondence.   For one sign of $\lambda$ various observables such as the deformed  energy spectrum have elegant gravity counterparts in terms of an AdS$_3$ geometry with a radial cutoff surface, with the position of the cutoff fixed by $\lambda$. The defining flow equation,   $\frac{dS_\lambda}{d\lambda } = \int d^2x\, \OTT$, can be reinterpreted as one component of the Einstein equations. See \cite{Kraus:2018xrn,Taylor:2018xcy,Hartman:2018tkw,Belin:2020oib,Kruthoff:2020hsi,Li:2020pwa} for some further developments and \cite{Giveon:2017myj,Guica:2019nzm,Hirano:2020nwq,Apolo:2019zai} for alternative interpretations.

The notion of a dynamical change of coordinates is very natural in the context of a holographic dual with a sliding cutoff surface \cite{McGough:2016lol}.   We will derive flow equations by demanding that the metric on the cutoff surface maintains its form, for instance by choosing the conformal gauge.  In the case of flat cutoff surfaces, the flow equations match those found by Cardy in his analysis of correlation functions \cite{Cardy:2019qao}, including the modification to the stress tensor flow required in order to maintain conservation.  A new feature is the appearance of a dynamical Weyl factor; in this context we remark that at the level of a classical bulk there is no obstacle to introducing a curved metric. Indeed, the complications mentioned above regarding the definition of the $\OTT$ operator in curved space have to do with a breakdown of factorization, but factorization is automatic in the classical bulk limit (\emph{i.e.} the large $N$ limit from the point of view of the boundary field theory).

There are several ways to motivate the relation between the \TT\ deformation and AdS$_3$ gravity with a radial cutoff.  We provide a new one which is very direct and illuminates previous treatments.   Specifically, in \cite{Guica:2019nzm} it was shown how the dictionary emerges by applying standard AdS/CFT rules to double trace interactions involving the stress tensor: adding the double trace interaction changes the variational principle, and the metric that should be held fixed in the variational principle was identified as the metric on the cutoff surface.  Here we proceed by  evaluating the on-shell action  of the region between the cutoff surface and the AdS$_3$ boundary.  This action turns out to be simply the double trace interaction $\int d^2x \OTT$ evaluated on the AdS$_3$ boundary.   From this it follows that including this interaction is equivalent to integrating out the region of the bulk exterior to the cutoff surface.  The holographic dictionary is  an immediate consequence.

We should note that the usefulness of a gravity dual for the \TT\ deformation becomes less clear for observables that are sensitive to bulk matter fields \cite{Kraus:2018xrn,Hartman:2018tkw,Guica:2019nzm}, and for this reason we will restrict attention to pure gravity in the bulk.  Our hope is that lessons learned in this context will turn out to have general validity.

The outline of this paper is as follows. In Section \ref{sec:TTb-deformation} we describe some general aspects of the $\ttb$ deformation in curved space, putting a special emphasis on the factorization property of the expectation value of the $T\bar{T}$ operator. The derivation of the dynamical change of coordinates and its application to computing deformed Lagrangians is the content of Section \ref{sec:TopoGravity}. We evaluate the annular action
in Section \ref{sec:holography1} and show its direct relation to classical on-shell actions on the field theory side. The dynamical coordinates are recovered from the cutoff AdS$_3$ setting in Section \ref{sec:dynCoordsAdS3}. In section \ref{sec:STflow} we derive the flow equation for the deformed stress tensor from holography. We conclude in Section \ref{sec:conclusions} and discuss potential future directions.

\section{The $\ttb$ deformation}
\label{sec:TTb-deformation}

Starting our discussion in flat space, a $\ttb$ deformed QFT is defined, at least formally,  in terms of a path integral with respect to a 1-parameter family of actions $S_\lambda$  that obeys the flow equation \cite{Smirnov:2016lqw,Cavaglia:2016oda}
\begin{align}\label{Sdef}
\frac{\rd S_{\lambda}}{\rd\lambda}=\int\! d^2 x\,\mathcal{O}_{T\bar{T}}^{(\lambda)}~,
\end{align}
where the $\OTT$ operator is quadratic in the stress tensor of the deformed theory,
\begin{align}
\label{eq:def1M}
\mathcal{O}_{T\bar{T}}=-\frac{1}{8}\epsilon_{\mu\nu}\epsilon_{\rho\sigma}T^{\mu\rho}T^{\nu\sigma},
\end{align}
and the initial condition $S_{\lambda=0}=S_0$ where $S_0$ is the  undeformed action.  More precisely, the composite $\OTT$  operator is defined by point splitting; as shown by \cite{Zamolodchikov:2004ce} the precise combination of stress tensors appearing in the definition implies that all divergences encountered in the coincident limit are proportional to total derivatives of local operators, and so  $\int\! d^2x \OTT$ is finite and unambiguous. We shall review in the next section an efficient algorithm for computing the deformed action $S_\lambda$.  As we will see, studying the classical deformed actions are interesting in their own right.\footnote{When it comes to analysing the quantum theory however, it is  far from clear whether the usual route of defining correlation functions non-perturbatively via the path integral makes sense, given the non-renormalizable nature of the theory.}

We now discuss what is known about the $\ttb$ deformation on a curved space with metric tensor $g_{\mu\nu}$.   As a first attempt we can introduce the metric in the obvious way by writing  $\frac{\rd S_{\lambda}}{\rd\lambda}=\int\! d^2 x \sqrt{g} \mathcal{O}_{T\bar{T}}^{(\lambda)}$ where now
\begin{align}
\label{eq:def1A}
\mathcal{O}_{T\bar{T}}=-\frac{g}{8}\epsilon_{\mu\nu}\epsilon_{\rho\sigma}T^{\mu\rho}T^{\nu\sigma}.
\end{align}
We denote the Levi-Civita symbol as $\eps_{\mu\nu} = \pm 1$ and write $g=|\det g_{\mu\nu}|$.  This $T\bar{T}$ operator has been considered in \cite{Bonelli:2018kik,Coleman:2019dvf,Tolley:2019nmm}. It can be written in several different ways (see \eg \cite{Caputa:2019pam,Jiang:2019tcq}). Using the identity
\begin{align}
g \eps_{\mu\nu}\eps_{\rho\sigma}=g_{\mu\rho}g_{\nu\sigma}-g_{\mu\sigma}g_{\nu\rho},
\end{align}
we can write
\begin{align}
\label{eq:def2}
\mathcal{O}_{T\bar{T}}=
\frac{1}{8}(T_{\mu\nu}T^{\mu\nu}-T_{\mu}^{\mu}T_{\nu}^{\nu}),
\end{align}
as well as
\begin{align}
\label{eqdet}
\mathcal{O}_{T\bar{T}}=-{1\over 4} \det T^\mu_\nu~.
\end{align}

It will also be useful to work in  the first order formalism. Let us introduce the zweibein $e_{\mu}^a$ such that
\begin{align}
g_{\mu\nu}=\delta_{ab} e_{\mu}^a e_{\nu}^b,
\end{align}
where the frame bundle metric is flat. The stress tensor with flat indices reads
\begin{align}
T^{ab}=e_{\mu}^a e_{\nu}^b T^{\mu\nu},
\end{align}
and we have
\begin{align}
\label{eq:def1B}
\mathcal{O}_{T\bar{T}}=-\frac{1}{8}\epsilon_{ab}\epsilon_{cd}T^{ac}T^{bd}~.
\end{align}

As we discuss in the next section, given an action $S_\lambda$ that satisifies  the flat space flow equation (\ref{Sdef}) it is simple to write down the curved space version.  However, the definition of $\OTT$ as a quantum operator is much more difficult.  One important property of the $\mathcal{O}_{T\bar{T}}$ operator in flat spacetime is that its expectation value on the cylinder factorizes \cite{Zamolodchikov:2004ce},
\begin{align}
\epsilon_{\mu\nu}\epsilon_{\rho\sigma}\langle n|T^{\mu\rho}T^{\nu\sigma}|n\rangle=
\epsilon_{\mu\nu}\epsilon_{\rho\sigma}\langle n|T^{\mu\rho}|n\rangle\langle n|T^{\nu\sigma}|n\rangle,
\end{align}
where $|n\rangle$ is an energy eigenstate of the Hamiltonian on the cylinder. This property underlies the quantum solvability of the $T\bar{T}$ deformation. It is therefore natural to ask whether the factorization property holds in curved spacetime. This is a difficult question for a generic curved metric $g_{\mu\nu}$, but some results are known for  maximally symmetric ones such as the sphere. Consider the vacuum expectation value $\langle\mathcal{O}_{T\bar{T}}\rangle$. The factorization property of $\langle\mathcal{O}_{T\bar{T}}\rangle$ was studied in \cite{Jiang:2019tcq} where it was shown that factorization does not hold in the presence of non-zero curvature, except in the large $c$ limit.

\section{Topological gravity and dynamical coordinates}
\label{sec:TopoGravity}
In this section, we review how the $\ttb$ deformation of QFTs can be interpreted as coupling the theory to  topological 2d gravity. We will then use this formulation to derive the dynamical change of coordinates. A gravity theory similar to Jackiw-Teitelboim gravity  was employed in \cite{Dubovsky:2017cnj}  and \cite{Dubovsky:2018bmo} to compute $\ttb$ deformed S-matrices and the torus partition functions.  Our treatment in this section is at the level of classical field theory, with the goal being to arrive at a deformed action $S_\lambda$ that obeys the flow equation and generalize the formalism to curved spaces.

\subsection{General solution of the flow equation}\label{flowseq}
 We follow the approach of \cite{Tolley:2019nmm} (see also related discussions in \cite{Freidel:2008sh,Mazenc:2019cfg}). We start from an undeformed action $S_0[g_{\mu\nu},\psi]$ defined on the metric $g_{\mu\nu}$ and where $\psi$ denotes matter fields.    We then introduce a second metric $\gamma_{\mu\nu}$, which will turn out to be the metric on which the deformed theory lives, and the action
\begin{align}\label{action-init}
S_{\lambda}=S_{\text{grav}}[g_{\mu\nu},\gamma_{\mu\nu}]+S_0[g_{\mu\nu},\psi].
\end{align}
$S_0[g_{\mu\nu}]$ is taken to be arbitrary --- it need not be conformal --- although we will assume for simplicity that there is no dependence on derivatives of the metric.

It is convenient to introduce zweibeins
\bea
  g_{\mu\nu}=\delta_{ab}e^{a}_\mu e^b_\nu~,\quad \gamma_{\mu\nu}=\delta_{ab}f^{a}_\mu f^b_\nu~,
\eea
where we choose to work in Euclidean signature.
 The action of the topological gravity is then\footnote{The Levi-Civita symbols have the conventions $\epsilon_{01}=1$ and $\epsilon^{01}=1$. }
\begin{align}
\label{eq:Sgravity}
S_{\text{grav}}[e^a_\mu,f^a_\mu]={1\over 2\pi^2\lambda} \int d^2 x\, \epsilon^{\mu\nu} \epsilon_{ab} (e^a_\mu-f_\mu^a)(e_\nu^b - f_\nu^b)~.
\end{align}
The deformed action $S_\lambda[f^a_\mu,\psi]$ is obtained by extremizing (\ref{action-init}) with respect to $e^a_\mu$ and substituting back.   We now verify that this procedure leads to a solution of the flow equation.

The stress tensor of the deformed theory is obtained by varying with respect to $f_\mu^a$
\begin{align}\label{defomed-stress-tensor}
T^\mu_a \equiv\frac{2\pi}{\det(f^a_\mu)}\frac{\delta S_\lambda[e,f,\psi]}{\delta f_\mu^a}=- \frac{2}{\pi\lambda \det(f^a_\mu)} \epsilon^{\mu\nu}  \epsilon_{ab} (e_\nu^b -f_\nu^b)~.
\end{align}
The equations of motion obtained by varying  the action \eqref{action-init} with respect to $e^a_\mu$ are
\begin{align}\label{eeq}
 \frac{1}{\pi^2 \lambda } {\epsilon}^{\mu\nu}  \epsilon_{ab} ({e^*}_\nu^b -f_\nu^b) + \frac{\delta S_0[e,\psi]}{\delta {e^*}_\mu^a}=0,
\end{align}
where we use $*$ to denote the on-shell solution.  This  equation can be rewritten in terms of  the stress tensor of the undeformed theory (denoted by $T_0$) as
\begin{align}\label{undefomed-stress-tensor}
T_{0\mu}^a = - \frac{ 2}{\pi \lambda \det(e^{*a}_\mu)} \epsilon^{\mu\nu}  \epsilon_{ab} ({e^*}_\nu^b -f_\nu^b)~.%
\end{align}

We now compute $dS_\lambda/d\lambda$.   Although $e^{*a}_\mu$ depends on $\lambda$, since we are extremizing the action with respect to $e^a_\mu$ we only need to differentiate $S_\lambda$ with respect to its explicit $\lambda$ dependence, which gives
\bea
{dS_\lambda \over d\lambda} = -{1\over 2\pi^2 \lambda^2} \int d^2 x\, \epsilon^{\mu\nu} \epsilon_{ab} (e^{*a}_\mu-f_\mu^a)(e_\nu^{*b} - f_\nu^b)~.
\eea
Raising and lowering indices on the stress tensor as
$
T_\nu^b = f_\mu^b f_\nu^a T^\mu_a,
$
we find from (\ref{defomed-stress-tensor}) the relation
\bea\label{detT}
\det (T^a_\mu) = \left( {2 \over \pi \lambda}\right)^2 {1\over 2}  \epsilon^{\mu\nu} \epsilon_{ab} (e^{*a}_\mu-f_\mu^a)(e_\nu^{*b} - f_\nu^b)~.
\eea
The change in the action is given by
\begin{align}\label{Sgravc}
{dS_\lambda \over d\lambda}  =      -{1\over 4} \int\! d^2x \det(T^a_\mu)  =  \int\! d^2x \sqrt{g} \,\OTT~,
\end{align}
which is our desired flow equation.  We also note that the on-shell value of $S_{\rm grav}$ is
\bea\label{Sgrav}
S_{\rm grav}[e^{*a}_\mu,f^a_\mu] = {\lambda \over 4} \int\! d^2x \sqrt{g} \det(T^\mu_\nu)~.
\eea

Finally, we can  compute the trace of the deformed stress tensor.   From (\ref{defomed-stress-tensor}) we have
\begin{align}\label{TrT}
T^\mu_\mu & = f^a_\mu T^\mu_a = - \frac{2}{\pi\lambda \det(f^a_\mu)} \epsilon^{\mu\nu}  \epsilon_{ab} f^a_\mu (e_\nu^{*b} -f_\nu^b) \cr
& = T^\mu_{0\mu}+  \pi \lambda \det (T^\mu_\nu) ,
\end{align}
where we used  (\ref{detT}) and (\ref{undefomed-stress-tensor}) to obtain the result in the second line.

Although the action  $S_{\text{grav}}$  considered here is a purely 2d action, we will see later that the on-shell value (\ref{Sgrav}) coincides with the on-shell action of  3d Einstein gravity in the  region contained between a cutoff surface and the AdS$_3$ boundary.

\subsection{Dynamical coordinates and dynamical Weyl transformation}
 Equations (\ref{defomed-stress-tensor}) and (\ref{undefomed-stress-tensor}) for the stress tensors may be recast as follows
\begin{align}\label{Tforms}
	e^a_\mu &= f_\mu^a - {\pi \lambda \over 2} \epsilon_{\mu\nu} \epsilon^{ab} T_b^\nu ~,\cr
	f^a_\mu &= e_\mu^{*a} + {\pi \lambda \over 2} \epsilon_{\mu\nu} \epsilon^{ab} T_{0b}^\nu~.
\end{align}
These relations between the zweibeins describe how the  geometry  changes in response to the stress tensor.
It is clear that the above construction suffers no significant complications when the metric is taken to be curved as compared to being flat.  More precisely, this statement relies on our starting  assumption that the undeformed action does not depend on derivatives of the metric.

To facilitate comparison with other results, it is worthwhile to revert momentarily to the second order/metric formalism to analyse the implications of the dynamical change of coordinates when the undeformed theory lives in curved space.
We start with (\ref{Tforms}), in particular take $e^{*a}_\mu$ and $T^\nu_{0b}$ as fixed (independent of $\lambda$). Differentiating the second equation gives
\bea
\p_\lambda f^a_\mu = {\pi \over 2} \eps_{\mu\nu}\eps^{ab}T^\nu_{0b}~.
\eea
Now, on-shell the equivalence of the two equations in  (\ref{Tforms}) implies $ \eps_{\mu\nu}\eps^{ab}T^\nu_{0b} = \eps_{\mu\nu}\eps^{ab}T^\nu_{b} $. Therefore we have the flow equation %
\bea
\p_\lambda f^a_\mu = {\pi \over 2} \eps_{\mu\nu}\eps^{ab}T^\nu_{b}~.
\eea
Let us denote the deformed metric as $\gamma_{\mu\nu} =\delta_{ab}f^a_\mu f^b_\nu$.  We can then evaluate the flow of this deformed metric using the flow of the zweibein above
\begin{align}\label{metricFlow}
	\p_\lambda \gamma_{\mu\nu} ~=~\delta_{ab}\p_\lambda f^a_\mu f^b_\nu+  \delta_{ab} f^a_\mu \p_\lambda f^b_\nu ~ =  ~{\pi \over 2}\delta_{ab} \left(\eps_{\mu \gamma}\eps^{ac} T^\gamma_c f^b_\nu + \eps_{\nu \gamma}\eps^{bc} f^a_\mu T^\gamma_c  \right)\,= \,-\pi \hat{T}_{\mu\nu}
\end{align}
where we have defined $\hat{T}_{\mu\nu} = T_{\mu\nu} - T^\alpha_\alpha \gamma_{\mu\nu}  = -{\eps}_{\mu}^{~\gamma}{\eps}_{\nu}^{~\sigma} T_{\gamma\sigma}$. In Section \ref{sec:holography1}, it will be shown that this flow equation will have a precise incarnation in holography.

The change of the metric \eqref{metricFlow} induces a change in the Weyl factor. We can write the metric in the conformal gauge as, $\gamma_{\mu\nu}dx^\mu dx^\nu = e^{2\omega(z,\bz)}dzd\bz$. The net change of metric due to infinitesimal diffeormorphisms, $\delta x^\mu = \epsilon^\mu$, and infinitesimal Weyl transformations, $\delta\omega(x)=\sigma(x)$, is
\begin{align}
	\delta  \gamma _{\alpha\beta} = 2\sigma \gamma_{\alpha\beta} + \nabla_{(\alpha} \epsilon_{\beta)}~.
\end{align}
Taking the trace of the above equation and using \eqref{metricFlow} for $\delta  \gamma _{\alpha\beta}$, we can extract the infinitesimal change in the Weyl factor
\begin{align}\label{dynamicalWeylFT}
	\sigma =-e^{-2\omega} (\p_z \eps_{\zb}+\p_{\zb}\eps_z)  - \frac{\pi}{4} T^i_i \delta \lambda~.
\end{align}
Here, we used \eqref{conformalGaugeFormulas} to rewrite the covariant derivatives.
This shows that the Weyl factor also becomes dynamical, as it depends on the fields of the theory through the deformed stress tensor.

Let us now discuss some aspects of the dynamical change of coordinates in the case when the deformed metric is flat, \ie $f_\mu^a=\delta_\mu^a$.  This implies that undeformed metric is also flat on-shell.  To see this we consider (\ref{Tforms}),
\begin{align}\label{dcc-flat}
e^a_\mu = \delta_\mu^a - {\pi \lambda \over 2} \epsilon_{\mu\nu} \epsilon^{ab} T_b^\nu.
\end{align}
Flatness of $e^a_\mu$ is equivalent to the statement that we can write $e^a_\mu = \p_\mu X^a$, where $X^a$ are new coordinates that put the line element in the form $ds^2 = dX^a dX^a$.  Inserting $e^a_\mu = \p_\mu X^a$ in (\ref{dcc-flat}) the existence of the $X^a$ requires that the integrability condition $\pd_\mu\pd_\nu X^a = \pd_\nu\pd_\mu X^a $  be obeyed, which is easily seen to be equivalent to the conservation of the stress tensor.  We conclude that flatness of $f^a_\mu$ implies flatness of $e^a_\mu$, only on-shell, since the stress tensor is only conserved on-shell.
Hence, the $X^a$'s exist on-shell and the solution is
\begin{align}
X^a (x) = - {\pi \lambda \over 2} \int_{x_0}^x dx'^\mu \epsilon_{\mu\nu}\epsilon^{ab}\,  T_b^\nu(x') ~.
\end{align}
Differentiating the with respect to $\lambda$ gives the flow equation for the dynamical coordinates
\begin{align}\label{dcc-flow}
\pd_\lambda X^a(x) = - {\pi \over 2} \int_{x_0}^x dx'^\mu \epsilon_{\mu\nu}\epsilon^{ab}\,  T_b^\nu(x').
\end{align}
This change of coordinates will be derived from the cutoff AdS$_3$ setup. As we already mentioned, the above change of coordinates can be  straightforwardly extended for the case of curved space. This can be demonstrated by  deriving deformed actions, which we turn to next.

\subsection{Classical deformed action}

We now provide a concrete example to illustrate the topological gravity formulation of the $T\bar{T}$ deformation.  Before we go into the technical details, let us make some general comments. From the topological gravity action, there are actually two approaches to compute the deformed Lagrangian.\par

The first approach, which is also the most direct, is to fix the zweibein $f_{\mu}^a$ and then solve the saddle-point equation for $e_{\mu}^a$. Denoting the solution as ${e^*}_{\mu}^a$, the $T\bar{T}$-deformed action on the curved background described by $f_{\mu}^a$ is given by $S_\lambda[f,e^*]$. For flat spacetime  we simply take $f_{\mu}^a=\delta_{\mu}^a$.\par

An alternative approach, employed in \cite{Coleman:2019dvf}, uses dynamical coordinates. Let us first recall what happens for the case of flat spacetime. Instead of fixing $f_{\mu}^a$, one fixes $e_{\mu}^a=\delta_{\mu}^a$ and then views the saddle-point equation as a definition of the dynamical coordinate $Y^a$ such that $f_{\mu}^a=\partial_{\mu}Y^a$. Then one performs a change of coordinates from $x^{\mu}$ to $Y^a$ in the original action to obtain the deformed action in  terms of the new coordinate $Y^a$. However, there is an important subtlety in this method which deserves  clarification. In order to define the dynamical coordinate $Y^a$ by writing $f_{\mu}^a=\partial_{\mu}Y^a$, one needs to check that the integrability condition $\partial_{\mu}\partial_{\nu}Y^a=\partial_{\nu}\partial_{\mu}Y^a$ holds.  However, just as was found in the last section for $X^a$, this condition amounts to stress tensor conservation, which only holds when  the matter fields are \emph{on-shell}.  This sounds bad, since we are after an expression for the deformed action valid for general field configurations.

It turns out that this subtlety doesn't actually matter in practice. Suppose we do impose the on-shell condition. Then we will obtain an action which obeys the flow equation on-shell. However, this action will in fact also obey the flow equation off-shell.
For example, in the case of a scalar field theory the on-shell conditions relate second derivatives of fields to lower derivatives, but when computing the stress tensor and verifying the flow equation, we never encounter such second derivatives. Hence it doesn't matter whether we are on-shell or off-shell.

Related to this, in the procedure of \cite{Coleman:2019dvf}, we do not actually need to compute the coordinate explicitly, we only need its derivative in order to convert derivatives using the chain rule.  We can think of formally solving for the quantity $f_{\mu}^a=\partial_{\mu}Y^a$ without worrying about whether $Y^a$ exists or not. Our argument in the previous paragraph justifies this procedure. To summarize, we can think of this approach as a useful method for computing deformed actions, but it is important to keep in mind that we are not really doing a well-defined coordinate change at the off-shell level.

To further illustrate these points  we now derive the deformed action for a scalar theory with a generic potential using both methods. The undeformed action is
\begin{align}
S_0[e,\phi]=\int d^2x\det(e_{\mu}^a)\,\mathcal{L}_0=\int d^2x\,\det(e_{\mu}^a)\,\left(\frac{1}{2}\delta^{ab}e^{\mu}_ae^{\nu}_b\partial_{\mu}\phi\partial_{\nu}\phi+V(\phi)\right),
\end{align}
where $V(\phi)$ is a generic potential that is independent of $e_{\mu}^a$. We do not restrict to flat metrics. In the rest of this section it is convenient to define $\lam  \equiv \pi^2 \lambda$.

\paragraph{Method 1} We fix the zweibein in  conformal gauge: $f_{\mu}^a=e^{\omega(x)}\delta_{\mu}^a$. The saddle-point equation (\ref{eeq})
can be solved as
\begin{align}
{e^*}_t^1=\frac{e^\omega}{2U} \left[1 +\frac{1-2\lam  U \phi_t^2}{\sqrt{1-2\lam  UK}}\right],
~~
{e^*}_t^2={e^*}_x^1=-\frac{\lam\,e^{\omega}\phi_t\phi_x}{\sqrt{1-2\lam  U K}},~~
{e^*}_x^2=\frac{e^\omega}{2U} \left[1 +\frac{1-2\lam  U \phi_x^2}{\sqrt{1-2\lam UK}}\right].
\end{align}
where
\begin{align}
\phi_t\equiv e^{-\omega}\partial_t\phi,\qquad \phi_x\equiv e^{-\omega}\partial_x\phi,\qquad K\equiv  (\phi_t^2+\phi_x^2) ,\qquad U=(1+\lam V).
\end{align}
Plugging into the original action (\ref{action-init}) and (\ref{eq:Sgravity}), we obtain
\begin{align}
S_{\lam}[f,e^*,\phi]=\int d^2x \det(f_{\mu}^a)\,\mathcal{L}_{T\bar{T}}=\int d^2x\,e^{2\omega}\,\mathcal{L}_{T\bar{T}},
\end{align}
where
\begin{align}
\label{eq:LTTbar}
\mathcal{L}_{T\bar{T}}=\frac{1+2\lam V-\sqrt{1-2\lambda(1+\lam V)K}}{2\lambda(1+\lam V)}.\qquad %
\end{align}
This matches the deformed Lagrangian derived earlier in \cite{Bonelli:2018kik,Cavaglia:2016oda} using different approaches.

\paragraph{Method 2} To apply the second approach we first identify the dynamical coordinates. We start with the equation of motion (\ref{eeq})
and fix  $e_{\mu}^a$ in  conformal gauge $e_{\mu}^a=e^{\omega}\delta_{\mu}^a$. To define the dynamical coordinate we factor out the same conformal factor and write $f_{\mu}^a=e^{\omega}\partial_{\mu}X^a$. This leads to
\begin{align}
\partial_{\mu}X^a=\delta_{\mu}^a+{\lam \over 2\pi } \,e^{-\omega}\,\epsilon_{\mu\nu}\epsilon^{ab}\det(e_{\nu}^b)T_{0b}^{\nu}~.
\end{align}
We rewrite the derivative of the scalar field by the chain rule
\begin{align}\label{dphi}
\partial_{\mu}\phi=\partial_{\mu}X^a\partial_a\phi~.
\end{align}
Plugging in the definition for the dynamical coordinate, we find $\partial_{\mu}X^a$ in terms of fundamental fields. The solution is
\begin{align}
\partial_t X^1=&\,\frac{\phi_T^2-(1-2\lam UK)\phi_X^2-(\phi_T^2-\phi_X^2)\sqrt{1-2\lam UK}}{\lam K^2},\\\nonumber
\partial_t X^2=&\,\partial_x X^1=\frac{2\phi_T\phi_X\left(1-\lambda UK-\sqrt{1-2\lam UK}\right)}{\lam K^2},\\\nonumber
\partial_x X^2=&\,\frac{\phi_X^2-(1-2\lam UK)\phi_T^2-(\phi_X^2-\phi_T^2)\sqrt{1-2\lam UK}}{\lam K^2},
\end{align}
where
\begin{align}
\phi_T=e^{-\omega}\partial_{X^1}\phi,\qquad \phi_X=e^{-\omega}\partial_{X^2}\phi,\qquad K=\phi_T^2+\phi_X^2,\qquad U=1+\lam V.
\end{align}
Computing $\p_\mu \phi$ from (\ref{dphi}) and inserting this  into (\ref{action-init}), we obtain the same deformed Lagrangian as in (\ref{eq:LTTbar}). This generalizes the approach of \cite{Coleman:2019dvf} to curved spaces.

\section{$T\bar{T}$ in curved space versus cutoff holography}
\label{sec:holography1}
In the following section we will offer a holographic perspective on the $T\bar{T}$ deformations in curved space using gravity in AdS$_3$. Within the framework of finite cutoff holography \cite{McGough:2016lol}, we will argue that there is a simple way to understand the bulk realization of $T\bar{T}$ deformation and its relation to the field theory constructions, shedding light on previous treatments  \cite{McGough:2016lol,Kraus:2018xrn,Guica:2019nzm,Hirano:2020nwq}.

First we establish that if we impose a radial cutoff and use the bulk action inside the cutoff geometry, then the stress tensor defined on that surface obeys the trace flow equation (which includes the Ricci scalar term).  We then write the cutoff AdS action as the full AdS action minus the action on the annular region between the cutoff and the AdS boundary. We will show that  this way of defining the cutoff action matches the classical field theory results in previous sections (the match extends to quantum level if ones uses the CFT Weyl anomaly in (\ref{TrT}), $T^\mu_{0\mu}= -{c\over 12}R$  ).

\subsection{3d gravity in AdS}
 We start by collecting relevant formulae regarding 3d gravity with a negative cosmological constant. We follow the convention of \cite{Kraus:2018xrn}.\footnote{Henceforth, we shall use Latin indices to denote curved coordinates. These are not to be confused with the frame bundle coordinates of Section  \ref{sec:TTb-deformation} and \ref{sec:TopoGravity}.}    The action, in Euclidean signature, is
\be\label{Einac}
S = -{1\over 16\pi G} \int_M\,d^3x \sqrt{g}(R+2) -{1\over 8\pi G}\int_{\partial M}\,d^2x \sqrt{h} (  K-1)+S_{\rm anom}~.
\ee
We have set the AdS radius to $1$. The Ricci scalars are $R(AdS_3)=-6$ and $R(S^2)=2$ in our convention.  $K$ is the trace of the extrinsic curvature on the boundary.   The boundary term $S_{\rm anom}$, needed for cancellation of divergences associated with the Weyl anomaly, is coordinate dependent and will be written below after fixing our coordinates. We employ the Fefferman-Graham coordinates
\be\
ds^2 = \frac{d\rho^2}{4\rho^2}+ h_{ij}(x,\rho)dx^i dx^j=  \frac{d\rho^2 }{ 4\rho^2} + \frac{1}{\rho} \gamma_{ij}(x,\rho) dx^i dx^j.\label{FGmet}
\ee
The AdS$_3$ boundary lies at $\rho=0$ and $\gamma_{ij}(x,0)$ is the metric on which the dual CFT lives.    In these coordinates the extrinsic curvature tensor for a surface of constant $\rho$ is
\be
K_{ij}=-\rho \partial_\rho h_{ij}\label{Kij}~,
\ee
and
\be
S_{\rm anom} = -{\ln (\rho_c) \over 32 \pi G} \int_{\p M} d^2x \sqrt{h}\,R(h)~.
\ee
$S_{\rm anom}$ depends explicitly on the {\em coordinate} location $\rho_c$ of the boundary, and hence is not diffeomorphism invariant. This is a manifestation of the Weyl anomaly in gravity language \cite{Henningson:1998gx,deHaro:2000vlm}.

We place the cutoff surface $\p M$ at $\rho=\rho_c$.  The asymptotic AdS boundary corresponds to $\rho_c\rt 0$.    The radial coordinate will lie in the interval $\rho_c \leq \rho \leq \rho_+$. The value of $\rho_+$  could arise either from a smooth endpoint of the geometry, as at a Euclidean black hole horizon, or from a breakdown of the Fefferman-Graham coordinates.  Its precise value will not play any role in what follows. We emphasize that we do not treat $\rho=\rho_+$ as part of $\p M$ in the sense that in (\ref{Einac}) there is no boundary term at $\rho=\rho_+$.  We rewrite the action using the identity
\bea
\sqrt{g}(R+2) = 2\p_\rho (\sqrt{h}K)+ {1\over 2\rho} \sqrt{h}\big(K^2-K^{ij}K_{ij}+2+R(h)\big)~,
\eea
which gives
\begin{align}
S &= -{1\over 16\pi G}\int^{\rho_+}_{\rho_c}{d\rho\over 2\rho}   \int\! d^2x \sqrt{h} \left( K^2 - K^{ij}K_{ij} +2+R(h)\right) \cr
 & \quad  +{1\over 8\pi G}\int\! d^2x \sqrt{h(\rho_c)}-{1\over 8\pi G}\int\! d^2x \sqrt{h(\rho_+)}K(\rho_+) +S_{\rm anom}.  \label{OSA}
\end{align}
In this subsection we  raise indices with $h^{ij}$,  where $h^{ik}h_{kj}=\delta^i_j$.\par
Einstein's equations can then be written as
\bea
&& K^2- K^{ij}K_{ij}=R(h)+2\,,\label{EE1}\\
&& \nabla^i (K_{ij}-K h_{ij}) =0\,, \label{EE2}\\
&&2\rho \partial_\rho( K_{ij}-h_{ij} K) +2 K_{ik}K^k_j- 3KK_{ij}+{1\over 2}h_{ij} \left[ K^{mn}K_{mn}+K^2 \right] - h_{ij}=0 \,,\label{EE3}
\eea
where $\nabla_i$ is the covariant derivative with respect to $h_{ij}$.

The general solution of Einstein's equations can be written as \cite{Skenderis:1999nb}
\bea\label{gensol}
 h_{ij}=\frac{1}{\rho}\gamma_{ij} = \frac{1}{\rho} \left(g^{(0)}_{ij} + \rho g^{(2)}_{ij} + {\rho^2\over 4} (g^{(2)} g_{(0)}^{-1} g^{(2)})_{ij}\right),
  \label{hGen}
\eea
where $g^{(0)}_{ij}$ and $g^{(2)}_{ij}$ are  $\rho$-independent.   For a given choice of $g^{(0)}_{ij}$, Einstein's equations fix the trace and divergence of $g^{(2)}_{ij}$. The trace condition is
\be
\tr(g^{-1}_{(0)}g^{(2)})=-\frac{1}{2}R(g^{(0)}).\label{Trg2}
\ee
The divergence condition is equivalent to conservation of the boundary stress tensor, defined momentarily.  Going back to \eqref{OSA}, we can read off the boundary stress tensor by varying the on-shell action with respect to  $h_{ij}$ \cite{Balasubramanian:1999re}
\be\label{Tdef}
 \delta S = \frac{1}{4\pi} \int_{\partial M} d^2x \sqrt{h} T^{ij}\delta h_{ij} = -\frac{1}{4\pi} \int_{\partial M} d^2x \sqrt{h} T_{ij}\delta h^{ij}.
 \ee
It works out to be
\be
T_{ij} = \frac{1}{4 G} (K_{ij}- K h_{ij} + h_{ij} ).\label{TijKij}
\ee
Taking its trace,
\be
 T^i_i = \frac{1}{4G}(2-K),
 \ee
and using  equation \eqref{EE1} we arrive at the following identity
\be
 T^i_i = 4G \det T^i_j - {1\over 8G} R(h)~.\label{TiiGr}
\ee
The $\rho\to 0$ limit yields the CFT Weyl anomaly
\be\label{Weyl-anom}
 (g_0^{-1})^{ij} T_{ij} = -{c\over 12} R(g_0)~,  \quad  c= {3\over 2G}~.
  \ee

For later purposes, it will be useful to write the stress tensor explicitly in terms of the solution \eqref{hGen}. For this, it is most convenient to introduce
\be
 \hat{T}_{ij} \equiv T_{ij} - T^k_k h_{ij},\label{That}
 \ee
for which
\be
\hat{T}_{ij}  =\frac{1}{4G}(K_{ij}-h_{ij}) = -\frac{1}{4G} \left( g^{(2)}_{ij} +\frac{1}{2}\rho( g^{(2)}g_{(0)}^{-1} g^{(2)})_{ij} \right). \label{Thatg}
\ee
Finally, the  stress tensor at the AdS boundary is obtained by taking the $\rho\rt 0$ limit
\be\label{Tcft}
T^{(0)}_{ij} = -\frac{1}{4G} \left(g^{(2)}_{ij} - \tr(g_{(0)}^{-1} g^{(2)})\, g^{(0)}_{ij}\right).
\ee
\subsection{Connection to $\ttb$ deformation}
Everything up to this point was a review of known facts about pure gravity in AdS$_3$ with a finite cutoff. Let us now briefly discuss the connection to the $T\bar{T}$ deformation of holographic CFTs.

Referring to (\ref{FGmet}) the deformed theory is thought of as living on  the $\rho=\rho_c$ surface with metric $\gamma_{ij}$.   In this section we will use $\gamma^{ij}$ to raise indices, where $\gamma^{ik}\gamma_{kj}=\delta^i_j$.  It is easy to check that the boundary stress tensor $T_{ij}$ with lower indices is unchanged whether we use (\ref{Tdef}) or $\delta S = -{1\over 4\pi} \int_{\p M} \! d^2x \sqrt{\gamma} T_{ij}\delta \gamma^{ij}$.  However, (\ref{TiiGr}) now takes a slightly different form
\be
 T^i_i = 4G \rho_c \det T^i_j - {1\over 8G} R(\gamma)~.\label{TrT}
\ee
Comparing to the trace flow equation in $T\bar{T}$ deformed CFT (following conventions from \cite{Kraus:2018xrn})
\be
 T^i_i = \pi \lambda \det T^i_j ,
\ee
 we see that we get an agreement if we identify
\bea\label{lamval}
\lambda = {4G \rho_c \over \pi}
\eea
and use the CFT Weyl anomaly \eqref{Weyl-anom}. This minimal treatment of the holographic dictionary for $T\bar{T}$ deformation will be sufficient for what follows, the reader can refer to \cite{McGough:2016lol,Kraus:2018xrn} for further discussions. %

\subsection{Annular action}
In order to further elucidate the bulk realization of the $\ttb$ deformation,  we compute the on-shell bulk action for the region between the cutoff surface at $\rho=\rho_c$ and the AdS$_3$ boundary at $\rho=0$.   We refer to this as the annular action since the constant time slices are annuli, at least for global AdS.   This computation will reveal a close connection to $S_{\rm grav}$ in (\ref{action-init}).\footnote{P.K. thanks Ben Michel for discussions and collaboration on this topic.  }

Before turning to the computation there is one issue worth clarifying.  We might imagine formulating a boundary value problem in which we solve Einstein's equations with specified metrics on the two boundaries.  However, it is important to note that Einstein's constraint equations do not allow us to freely specify the two boundary metrics independently.  This is most easily seen from  (\ref{gensol}).  Once the AdS boundary metric $g^{(0)}_{ij}$ is fixed, $g^{(2)}_{ij}$ is subject the trace and divergence conditions, and there is not enough freedom to specify the metric on the $\rho=\rho_c$ surface.

Keeping this in mind, we proceed.  $\p M$ now has two components, at $\rho=\rho_c$, and at the (regulated) AdS boundary $\rho=\eps$.  We include the boundary terms ${1\over 8\pi G} \int\! d^2x \sqrt{h}(K-1)$, with opposite signs, on the two boundaries.  Taking this  into account, we define the annular action as
\begin{align}
S_{\rm ann} & = -{1\over 16\pi G}\int^{\rho_c}_{\eps}{d\rho\over 2\rho}   \int\! d^2x \sqrt{h} \left( K^2 - K^{ij}K_{ij} +2+R(h)\right) \cr
 & \quad +{1\over 8\pi G}\int\! d^2x \sqrt{h(\eps)}-{1\over 8\pi G}\int\! d^2x \sqrt{h(\rho_c)}+S_{\rm anom},  \label{Sann}
\end{align}
where we take $\eps \rt 0$ at the end of the computation. The anomaly term is
\be
S_{\rm anom} = {\ln (\rho_c/\eps) \over 32 \pi G} \int\! d^2x \sqrt{h}R(h)~.
\ee
We have defined $S_{\rm ann}$ such that the total action for the region $\eps\leq \rho \leq \rho_+$ is given by the sum of (\ref{OSA}) and $S_{\rm ann}$, and consistent with this we see that $S_{\rm ann} \rt 0$ as $\rho_c \rt \epsilon$.
Using Einstein's equations we can write
\begin{align}
S_{\rm ann} & = -{1\over 4\pi G}\int^{\rho_c}_{\eps}{d\rho\over 2\rho}   \int\! d^2x \sqrt{h} +{1\over 8\pi G}\int\! d^2x \sqrt{h(\eps)}-{1\over 8\pi G}\int\! d^2x \sqrt{h(\rho_c)}-S_{\rm anom} .  \label{Sann2}
\end{align}
From the general  solution for $h_{ij}$ and identities in appendix \ref{ApA}, we can write
\be
\sqrt{h} =\sqrt{\det(g^{(0)})}    \left(\frac{1}{\rho } -  \frac{1}{4} R(g^{(0)}) +\frac{1}{4}\rho \det( g_{(0)}^{-1}g^{(2)} ) \right)~. \label{sqrth}
\ee
Using this and taking the $\eps \to 0$ limit gives
\be
S_{\rm ann}= -{\rho_c\over 16\pi G}  \int\! d^2x \sqrt{g^{(0)}} \det(g_{(0)}^{-1}g^{(2)})~.
\ee

The above result can be rewritten in various ways. From the expression (\ref{Tcft}) for the asymptotic AdS stress tensor (identified with the dual CFT stress tensor) we have
\be
 \det(g_{(0)}^{ik} T^{(0)}_{kj}) = \frac{1}{16 G^2} \det (g_{(0)}^{-1} g^{(2)})~.
\ee
We therefore have
\begin{align}
S_{\rm ann}& =\lambda \int\! d^2x \sqrt{g^{(0)}} \OTT^{(0)} ~, \label{Sat}
\end{align}
where we used $\lambda = 4G \rho_c/\pi$ and wrote
\bea
\OTT^{(0)} = -{1\over 4}  \det(g_{(0)}^{ik} T^{(0)}_{kj})~.
\eea
Alternatively, we can  use \eqref{hGen} and \eqref{That} to write
\be
S_{\rm ann} %
=      \lambda \int\, d^2x \sqrt{\gamma} \OTT ,\label{SaT}
\ee
where
$
\OTT = -{1\over 4}  \det(\gamma^{ik} T_{kj}).
$
The expressions \eqref{Sat} and \eqref{SaT} are the main results of this section.

\subsubsection{Example: global AdS$_3$}

Starting from AdS$_3$ in  standard global coordinates, $ds^2 = {dr^2 \over 1+r^2}+(1+r^2)dt^2 +r^2 d\phi^2$ we pass to Fefferman-Graham coordinates by writing
\bea
r= {4-\rho \over 4 \sqrt{\rho} }
\eea
which brings the metric to the form
\bea
ds^2 = {d\rho^2 \over 4\rho^2} + {1\over \rho}(dt^2+ d\phi^2) + {1\over 2}(dt^2 -d\phi^2) + {\rho \over 16}(dt^2 +d\phi^2)~,
\eea
so that $g^{(0)}_{ij} dx^i dx^j = dt^2+d\phi^2$ and $g^{(2)}_{ij}dx^i dx^j = {1\over 2}(dt^2 -d\phi^2)$.   From (\ref{Tcft}) the AdS$_3$ boundary stress tensor has nonzero components
\bea
T^{(0)}_{tt}= -{1\over 8G}~,\quad T^{(0)}_{\phi\phi}= {1\over 8G}~.
\eea
In global AdS$_3$  we have the option of shrinking the cutoff surface to zero, so that the annular region comprises all of AdS.  The origin is at $r=0$ which translates to $\rho_c=4$, which from (\ref{lamval}) translates into $\lambda = 16G/\pi$.  However, the annular action with inner surface at $\rho_c=4$ differs from the standard action we assign to global AdS, as we now discuss.

A straightforward computation of the annular action with $\rho_c =4$ yields $S_{\rm ann}= {1\over 8G} \Delta t$, where we consider a finite coordinate time interval of duration $\Delta t$.   We also have $\OTT^{(0)} = -{1\over 4} T^{(0)}_{tt} T^{(0)}_{\phi\phi} = {1\over 256 G^2}$.    This gives
\bea
\lambda \int\! d^2x \sqrt{g^{(0)}} \OTT^{(0)} = \left({16G\over \pi}\right)\left({1\over 256 G^2}\right) \left(2 \pi \Delta t\right) = {1\over 8G}\Delta t~,
\eea
which indeed reproduces the value of the annular  action.

On the other hand, if we start from (\ref{Einac}) and compute the action for global AdS we find $S_{\rm AdS} = -{1\over 8G}\Delta t=-{c\over 12}\Delta t$.  This result is consistent with the statement that the ground state energy of a CFT on the cylinder is $E_0 = - {c\over 12}$.

$S_{\rm ann}$ and $S_{\rm AdS}$ therefore differ by a sign, so that $S_{\rm ann}-S_{\rm AdS} = {1\over 4G}\Delta t$.    The reason for the difference is that $S_{\rm ann}$ includes inner boundary terms, while these are absent for $S_{\rm AdS}$.  Even though the inner boundary shrinks to zero size as $\rho_c\rt 4$, the Gibbons-Hawking term contributes in the limit.   Indeed  $\sqrt{h} K = {16+\rho^2 \over 8\rho}$ which is finite at $\rho=4 $ and gives
\bea S_{GH}= {1\over 8\pi G} \int_{\rho=4} d^2x \sqrt{h} K = {1\over 4 G}\Delta t~.
\eea
We see that this accounts for $S_{\rm ann}-S_{\rm AdS} $.

\subsection{Annular action and holographic $\ttb$ dictionary}
The results of the last subsection provide an illuminating way to think about the bulk dual of $\ttb$ deformation.  The main result is that the on-shell bulk action for the annular region is equal to the $\OTT$ operator integrated over either the inner or the outer boundary, where the $\OTT$ operator should be built out of the metric and stress tensor on the corresponding boundary.

Let us define $S_{\rm AdS}$ to be the action for the full AdS space with no cutoff surface.  We also define $S_\lambda$ to be the bulk action for the region contained within the cutoff surface.   The result of the last section can be expressed as
\bea
S_\lambda = S_{\rm AdS} -S_{\rm ann} = S_{\rm AdS} - \lambda \int\! d^2x \sqrt{g^{(0)}} \OTT^{(0)}~.
\eea
We immediately see the connection to the field theory discussion in Section \ref{flowseq}  where we wrote $S_\lambda = S_0 + S_{\rm grav}$.  Recall that  $S_0$ is the CFT action, which corresponds to $S_{\rm AdS}$, it follows from (\ref{Sgravc}) that $S_{\rm grav} = -S_{\rm ann}$.    The two constructions are therefore equivalent on-shell. There is no simple correspondence off-shell since the functional defining $S_{\rm grav}$ is very different from the 3d Einstein-Hilbert action.

This also provides us a useful perspective on the construction of \cite{Guica:2019nzm}, in which they arrived at the holographic dictionary by applying the standard rules of AdS/CFT in the presence of double-trace interactions.  Given the above results, we can immediately write
\bea
\delta \left(  S_{\rm AdS} - \lambda \int\! d^2x \sqrt{g^{(0)}} \OTT^{(0)} \right) = {1\over 4\pi} \int\!d^2x \sqrt{\gamma} T^{ij}\delta\gamma_{ij}~,
\eea
where the ``deformed stress tensor" $T^{ij}$ obeys the defining trace relation (\ref{TrT}) of a $\ttb$ deformed CFT.   In our construction the double trace interaction is identified with the annular action.

An interesting feature is that the annular action is linear in $\lambda$ when expressed in terms of the AdS boundary stress tensor, as in (\ref{SaT}).  At first glance, this appears in conflict with the statement that the $\ttb$ deformation modifies the action nonlinearly in $\lambda$.   The point is that this nonlinearity occurs when we perform the dynamical change of coordinates, which from the bulk point of view means working on the cutoff surface.  This linearization with respect to $\lambda$ is one of the main virtues of formulating the theory in terms of two metrics.

We should also emphasize that our simple result for the annular action is a reflection of the fact that pure 3d gravity has no local degrees of freedom. If we add propagating matters to the theory, there would be no way to write the action in terms of local expressions on the boundary.  For this reason, including propagating matters in the bulk will be much more complicated.

\section{Dynamical coordinates from gravity }
\label{sec:dynCoordsAdS3}

 In this section we derive the dynamical change of coordinates and dynamical Weyl transformation  from gravity.  The picture is very simple.  Given \eqref{FGmet}, the metric on a given fixed $\rho$ slice  can be brought into the flat form by a combination of a coordinate  and Weyl transformation.  These transformations will depend on $g^{(2)}_{ij}$ and are thus ``stress tensor dependent".

 In the following, we work this out in detail. We start with the bulk metric \eqref{FGmet} with fixed $g^{(0)}_{ij}$. By definition, $g^{(0)}_{ij}$ is identified with the $\rho\to 0$ limit of $\gamma_{ij}$.  We recall that a change in the location of the cutoff surface is related to a change in the $\ttb$ coupling as
\be\label{couplingDictionary}
\delta \lambda = {4G\over \pi }\delta \rho~.
\ee
\subsection{Radial fluctuations of the cutoff surface}

Consider the variation of $\gamma_{ij}$  under a small change of $\rho$,
\be
\delta_\rho \gamma_{ij}= \left( \rho \p_\rho h_{ij} + h_{ij}\right)\delta \rho
 =  \left( -K_{ij} + h_{ij}\right)\delta \rho~.
\ee
Let's express this in terms of the stress tensor on the cutoff surface.  From \eqref{Thatg},
\be\label{g-rho}
\delta_\rho \gamma_{ij}  =-4G \hat{T}_{ij}\delta \rho~,
\ee
 with
\be
\hat{T}_{ij} = T_{ij} - (h^{mn}T_{mn})h_{ij} =  T_{ij} - (\gamma^{mn}T_{mn})\gamma_{ij}~.
\ee
Upon using the the identification \eqref{couplingDictionary}, the change of the metric \eqref{g-rho} matches exactly with the flow equation \eqref{metricFlow} derived from coupling the field theory to 2d gravity.
Under a coordinate transformation $\delta x^i = \eps^i(x)$ that does not involve $\rho$  we have
\be\label{g-eps}
\delta_\eps \gamma_{ij} =  \nabla_i \eps_j + \nabla_j \eps_i~.
\ee
Note that here $\eps_i = \gamma_{ij}\eps^j$.  By choosing $\eps^i$ appropriately we can keep $\gamma_{ij}$ fixed up to a possible Weyl factor;   \ie we can find an $\eps^i(x)$ such that  $\delta_\rho \gamma_{ij} + \delta_{\eps} \gamma_{ij} =-2 \sigma \gamma_{ij}$, for some infinitesimal Weyl factor $\sigma$.  Therefore from \eqref{g-rho} and \eqref{g-eps}, we obtain
\be
 \nabla_i \eps_j + \nabla_j \eps_i+ 2\sigma \gamma_{ij} = {4G} \hat{T}_{ij}\delta \rho.\label{EQ2}
\ee
Below we show how to solve this, first for the flat metric and then for a general curved metric. %
\subsection{Flat boundary metric}

We first consider the simple case of a flat boundary  metric $\gamma_{ij}=\delta_{ij}$, where no Weyl factor appears under radial evolution.
Equation \eqref{EQ2} becomes
\be
 \p_i \eps_j = \frac{\pi}{2} \hat{T}_{ij}\delta \lambda.
\ee
Using $\hat{T}_{ij} =-\eps_{ik}\eps_{jl}T^{kl}$ we have
\be
 \p_i \eps_j = -\frac{\pi}{ 2}\eps_{ik}\eps_{jl}T^{kl}\delta \lambda .\label{EqFinF}
 \ee
This agrees with Cardy's result in \cite{Cardy:2019qao} and is consistent with the flow equation \eqref{dcc-flow}.  We note that in the above equation \eqref{EqFinF} $T_{ij}$ is defined on the metric of the cutoff surface, which indeed corresponds to the deformed stress tensor appearing in \cite{Cardy:2019qao}.
\subsection{General cutoff metric}

We now consider the general situation.  We write (\ref{EQ2}) as
\be
\label{eq:rewriteEQ}
\nabla_i \eps_j + \sigma \gamma_{ij} = {2G} \hat{T}_{ij}\delta \rho~.
\ee
Taking the trace of this equation and using $\hat{T}^i_i = -T^i_i$ leads o
\be
 \sigma =-{1\over 2} \nabla_i \eps^i  -G T^k_k \delta \rho.\label{sigma}
 \ee
Plugging this back in (\ref{eq:rewriteEQ}) gives the  traceless part
\be
\nabla_i \eps_j -\frac{1}{2} \nabla_k \eps^k \gamma_{ij} = {2G} \left( T_{ij}-\frac{1}{2} T^k_k \gamma_{ij}\right)\delta \rho~.\label{EQEps}
\ee
We  choose to work in conformal gauge by taking
\be
  \gamma_{ij}dx^i dx^j = e^{2\omega(z,\zb)} dzd\zb.
\ee
Using the identities in  Appendix \ref{ApA},  we can write the conservation equation for the stress tensor as
\bea
\p_{\zb}T_{zz} + e^{2\omega} \p_z(e^{-2\omega} T_{z\zb}) =0,\qquad
  \p_{z}T_{\zb\zb} +e^{2\omega} \p_{\zb}(e^{-2\omega} T_{z\zb}) =0 .
\eea
The equations in \eqref{EQEps} become
\bea
\p_z \eps^{\zb}  = 4G e^{-2\omega} T_{zz}\delta \rho, \qquad
   \p_{\zb}\eps^{z} = 4G e^{-2\omega} T_{\zb\zb}\delta \rho,  \label{EpsGT}
\eea
which can be solved by
\bea
\label{eq:shiftE}
\eps^{\zb}(z,\zb) & =& 4G \delta \rho \int_{z_0}^z e^{-2\omega(z',\zb)} T_{zz}(z',\zb) dz'  \nn\\
\eps^{z}(z,\zb) & =& 4G \delta \rho \int_{\zb_0}^{\zb}  e^{-2\omega(z,\zb')} T_{\zb\zb }(z,\zb') d\zb'. \label{EPSi}
\eea
Finally, we can compute the Weyl transformation $\sigma$ from \eqref{sigma},
\be
\sigma =-e^{-2\omega} (\p_z \eps_{\zb}+\p_{\zb}\eps_z)  - G\delta \rho T^i_i .\label{sigmaR}
\ee
This is in perfect agreement with \eqref{dynamicalWeylFT} obtained on the field theory side.
As $\sigma$ corresponds to an infinitesimal change in $\omega$, we can write (\ref{sigmaR}) as a flow equation
\be
{\p \omega \over \p \rho} = e^{-2\omega} (\p_z \eps_{\zb}+\p_{\zb}\eps_z)  + G T^i_i~.
\ee
This is the dynamical Weyl transformation that depends on the trace of the stress tensor.
If we write the coordinate change as $x^i \rightarrow v^i(x)$ then  \eqref{EPSi} implies the flow equations
\bea
{\p v^{\zb}(z,\zb) \over \p \rho} & = &4G  \int_{z_0}^z e^{-2\omega(z',\zb)} T_{zz}(z',\zb) dz'  \nn\\
{\p   v^{z}(z,\zb)\over \p \rho}  & =& 4G  \int_{\zb_0}^{\zb}  e^{-2\omega(z,\zb')} T_{\bz \bz}(z,\zb') d\zb'~.
\eea
This describes the dynamical change of coordinates and  is consistent with the analysis on the field theory side.
Moreover, $T_{ij}$ also flows due to the coordinate transformation, which we analyse  in the next section.
Altogether, we have a set of coupled nonlinear equations for $\eps^i$ and $\omega$ which appear rather difficult to solve in general.

\section{Flow equation for the stress tensor from holography }
\label{sec:STflow}
In addition to the utility of deriving deformed Lagrangians, the dynamical coordinates also offer a route to computing correlation functions of the deformed theory \cite{Cardy:2019qao}. It was shown in \cite{Cardy:2019qao} that  the correlators of deformed operators are equivalent to those of the undeformed operators but in the new dynamical coordinates. This formalism can also be used to analyse how operators themselves flow under $\ttb$. In this section we focus on the stress tensor, which is universal and has a natural analogue in holography.
The deformation of the stress tensor is given by the following flow equation
\be
D_\lambda T^\lambda_{dc}(x) ={\pi\over 2}\epsilon^{ab}\epsilon^{ij}\int_x^X dx'_j T^\lambda_{ai}(x')\partial_b T^\lambda_{dc}(x)-{\pi \over 2}  T^{\lambda a}_{~~c}(x) T^\lambda _{da}(x) \label{kka}~.
\ee
Here, $D_\lambda$ denotes the infinitesimal difference from $\lambda$ to $\lambda+\delta\lambda$.
The first term arises from the dynamical change of coordinates.  The second term is a correction piece that is required to preserve conservation of the stress tensor along the flow.

Our goal now is to derive the above flow equation from gravity.  We consider a flat boundary metric, since this is assumed in \cite{Cardy:2019qao}.   There are two effects in the bulk that contribute to the flow of the stress tensor.  First, there is a change due to the physical motion of the cutoff surface and, second, there is a change due to the coordinate transformation needed to put the metric on the new cutoff surface in standard form $ds^2 =dzd\zb$. These two effects will combine together to give a result that matches \eqref{kka}.

 The first step is to rewrite Einstein's equation \eqref{EE3} in terms of the boundary stress tensor \eqref{TijKij}.    For convenience we will work at the point $\rho=1$, which involves no loss of generality due to our freedom to rescale $\rho$.  This procedure leads to the following equations
\bea
\p_\rho T_{zz} =-8G T_{z\zb}T_{zz}, \qquad
   \p_\rho T_{\zb\zb} =-8G T_{z\zb}T_{\zb\zb}, \qquad
 \p_\rho T_{z\zb} =4G\big( (T_{z\zb})^2 -3 T_{zz} T_{\zb\zb} \big) .  \label{Tprimes}
\eea
This gives us the flow of the stress tensor before we make any coordinate transformation to keep the  metric $\gamma_{ij}= \rho h_{ij}$ fixed.   To implement the latter we refer to the result in \eqref{EqFinF}.    Setting $\gamma_{ij}=\delta_{ij} $ in \eqref{EqFinF}, we have the following relation for the diffeomorphism vector field
\be\label{deps}
\partial_i \eps_j = -2G \epsilon_{ik}\epsilon_{jl}T^{kl} \delta \rho~.
\ee
From \eqref{EpsGT} we have in complex coordinates
\bea \label{EpTadr}
 \partial_z \epsilon^{\zb}   =  4G  T_{zz}\delta \rho, \qquad
   \partial_{\zb}\epsilon^{z}  =  4G T_{\zb\zb}\delta \rho, \qquad
   \p_i \eps^i = -8G T_{z\zb}\delta \rho
   ~.
   \eea
   Now differentiating \eqref{EpTadr} and using the conservation law we have
   \be\label{id2}
   \p_{\zb} \p_z \eps^{\zb} = 4G \p_{\zb} T_{zz}\delta \rho = -4G \p_z T_{z\zb}\delta \rho~.
   \ee
   These are the necessary  ingredients for determining the change of the stress tensor under diffeomorphisms.
 Under the coordinate transformation by $\eps^i$,  change in the stress tensor is given by the Lie derivative
\be
\delta T_{ij} = \mathcal{L}_\eps T_{ij} = \eps^k (\p_k T_{ij}) + (\p_i \eps^k )T_{kj} + (\p_j \eps^k )T_{ik}~.
\ee
Using \eqref{EpTadr} and \eqref{id2}, we have the following changes for the components
\bea
\delta T_{zz} =  \eps^k \p_k T_{zz}, \qquad
  \delta T_{\zb\zb} =  \eps^k \p_k T_{\zb\zb}, \qquad
  \delta T_{z\zb}  = \eps^{k} \p_{k} T_{z\zb}  -8G  (T_{z\zb})^2 +8G T_{zz} T_{\zb\zb}~ .\label{Tzs}
  \eea
  Finally, we are ready to combine the two effects --  the change due to fluctuation of the cutoff surface, equation \eqref{Tprimes}, and due the coordinate change, equation \eqref{Tzs}. The net change is given by
\be
\Delta T_{ij} ~= ~\delta T_{ij} + \p_\rho T_{ij}\delta \rho~=  ~ \eps^k \p_k T_{ij}- \frac{\pi}{2} T_{ik}T^k_{~j} \delta \lambda~,
\ee
 where, we used   $\pi \delta \lambda = 4G \delta \rho$.
This is the main result of this section and it matches precisely with \eqref{kka} by plugging in the explicit expressions for $\eps^k$ from (\ref{eq:shiftE}), or equivalently by using (\ref{deps}).  One can easily check that  the flow preserves conservation, which is what was imposed in \cite{Cardy:2019qao} by adding the second term above.  From the bulk perspective this conservation is built-in.

\section{Conclusions}
\label{sec:conclusions}

In this work we illuminated some aspects of  $\ttb$ deformed field theories and their holographic avatars in terms AdS with a finite radial cutoff. On the field theory side, we showed how the deformation can be formulated via a dynamical change of coordinates and generalized this analysis to the situation when the undeformed theory lives on a curved space. We also provided a more direct means of deriving deformed Lagrangians using this machinery. The holographic side of our story refines and adds  a number of elements to the cutoff AdS proposal. Firstly, the role of placing a radial cutoff was made manifest by showing that the action of the annular region between the cutoff and the old AdS boundary is given by the $\ttb$ operator (integrated over either of these boundaries). The dynamical change of coordinates were also shown to naturally arise in this holographic setup by analysing radial fluctuations of the cutoff surface. Finally, we also uncovered how the flow equation for the deformed stress tensor has an exact parallel in the bulk. This was achieved by studying the change of the holographic stress tensor due to variations of the cutoff surface and an extra coordinate transformation to keep the metric flat. On the whole, these precise gravitational manifestations of the field theory $\ttb$ flows sheds light on why the cutoff AdS setup works.

There are many important points to be better understood, which constitute interesting future directions. The analysis performed here was entirely classical and we focused mostly on on-shell physics. However, the real challenges in understanding $\ttb$ theories arise at the quantum level.   Our hope is that the geometrical structures that appear at the classical level will survive in some form in the quantum theory, but to develop this we need to study observables in the quantum regime. Concretely, it remains to be seen how correlation functions can be  reconstructed from the cutoff geometry, both order by order in the $\ttb$ deformation parameter  $\lambda$ expansion and non-perturbatively.  It is reasonable to hope that the flow equation for the stress tensor correlators can be reproduced from the bulk.  At leading order in  $\lambda$,  2-point and 3-point correlations of the stress tensor have been computed in the cutoff gravity setup, and from conformal perturbation in the deformed CFT \cite{Kraus:2018xrn}. These were shown to match  and can be understood from demanding stress tensor conservation and the trace relation $T^i_i =\pi \lambda \det T^i_j$. It would be interesting to study this at higher orders in $\lambda$.

As a parting comment, we cannot resist pointing out the similarity between   the deformation and QFTs on non-commutative geometries. This is particularly tempting from the dynamical coordinates point of view. As we have seen, turning on the $\ttb$  deformation is equivalent to putting the theory on stress-tensor dependent coordinates. Thinking of these as operators, these dynamical coordinates then fail to commute. This may provide a route to decipher the theory's non-local features. Some similarities with the deformed S-matrix have been pointed out a while ago in \cite{Grosse:2007vr,Dubovsky:2012wk}. It would be tantalizing to make a clear identification.

\section*{Acknowledgements}
P.C. thanks Onkar Parrikar for discussions and Shinji Hirano for
comments on the draft.
The work of P.C.~is supported by NAWA ``Polish Returns 2019" and NCN Sonata Bis 9 grants.
S.D.~is grateful to the Pauli Center for Theoretical Studies, Z\"urich for the support and warm hospitality during the spring of 2020 when this work was initiated. P.K. thanks Eric D'Hoker, Ben Michel and Ruben Monten for discussions.   P.K. is supported in part by the National Science Foundation under research grant PHY-1914412.

\appendix
\section*{Appendix}

\section{Some useful identities}\label{ApA}
\subsection*{Matrix identities}
Given an arbitrary $2\times 2$ matrix $M$ we define
\be
\hat{M} = M- \tr(M) I
\ee
and note the following relations between the traces, determinants and inverses
\begin{align}
& \tr(\hat{M}) = -\tr(M) , \qquad
\tr(\hat{M}^2) = \tr(M^2),\nn\\
&\det (\hat{M}) = \det (M) ,\qquad
M^{-1}= -{1\over \det M} \hat{M} ,\qquad
\hat{M}^{-1}  = - {1\over \det M} M.\nn
\end{align}
Also useful are
\bea
 \det(M) & =& {1\over 2} \big[ (\tr M)^2 - \tr(M^2)\big] \nn\\
\det(I+M) & =& 1 + \tr(M) + \det(M).
\eea
\subsection*{2d conformal gauge}
Some of the calculations in the main text were performed in conformal gauge metric in two dimensions
\be
ds^2=e^{2\omega}dzd\bar{z},
\ee
The following formulas were also used
\bea\label{conformalGaugeFormulas}
 &&R=-2e^{-2\omega} \p_i \p_i\omega=-8 e^{-2\omega} \p_z \p_{\zb}\omega ~,\nn\\
&&\Gamma^{z}_{zz} = 2\p_z \omega~,\quad \Gamma^{\zb}_{\zb\zb} = 2\p_{\zb} \omega~,\nn\\
&&  \nabla_z \eps_z =e^{2\omega}\p_z( e^{-2\omega} \eps_z)~,\quad \nabla_{\zb} \eps_{\zb} = e^{2\omega}\p_{\zb}( e^{-2\omega} \eps_{\zb})~,\quad \nabla_z \eps_{\zb} = \p_z \eps_{\zb}~,\quad \nabla_{\zb}\eps_z = \p_{\zb}\eps_z \nn\\
&& \nabla_i \eps^i = 2e^{-2\omega} (\p_z \eps_{\zb}+\p_{\zb}\eps_z)~.
\eea
The other Christoffel symbols vanish.

\providecommand{\href}[2]{#2}


\begingroup\begin{thebibliography}{10}
	
	\bibitem{Smirnov:2016lqw}
	F.~A. Smirnov and A.~B. Zamolodchikov, {\it {On space of integrable quantum
			field theories}},
	\href{http://dx.doi.org/10.1016/j.nuclphysb.2016.12.014}{{\sf Nucl. Phys.}
		{\sf {B915} }{\sf (2017) }{\sf 363--383}},
	\href{http://arxiv.org/abs/1608.05499}{{\ttfamily arXiv:1608.05499 [hep-th]}}.
	%
	
	\bibitem{Cavaglia:2016oda}
	A.~Cavagli{\`a}, S.~Negro, I.~M. Sz{\'e}cs{\'e}nyi, and R.~Tateo, {\it {$T
			\bar{T}$-deformed 2D Quantum Field Theories}},
	\href{http://dx.doi.org/10.1007/JHEP10(2016)112}{{\sf JHEP} {\sf {10} }{\sf
			(2016) }{\sf 112}},
	\href{http://arxiv.org/abs/1608.05534}{{\ttfamily arXiv:1608.05534 [hep-th]}}.
	%
	
	\bibitem{Jiang:2019hxb}
	Y.~Jiang, {\it {Lectures on solvable irrelevant deformations of 2d quantum
			field theory}},  \href{http://arxiv.org/abs/1904.13376}{{\ttfamily
			arXiv:1904.13376 [hep-th]}}.
	
	\bibitem{Giveon:2019fgr}
	A.~Giveon, {\it {Comments on $T\bar T$, $J\bar{T}$ and String Theory}},
	\href{http://arxiv.org/abs/1903.06883}{{\ttfamily arXiv:1903.06883
			[hep-th]}}.
	
	\bibitem{Dubovsky:2017cnj}
	S.~Dubovsky, V.~Gorbenko, and M.~Mirbabayi, {\it {Asymptotic fragility, near
			AdS$_{2}$ holography and $ T\overline{T} $}},
	\href{http://dx.doi.org/10.1007/JHEP09(2017)136}{{\sf JHEP} {\sf {09} }{\sf
			(2017) }{\sf 136}},
	\href{http://arxiv.org/abs/1706.06604}{{\ttfamily arXiv:1706.06604 [hep-th]}}.
	%
	
	\bibitem{Cardy:2018sdv}
	J.~Cardy, {\it {The $ T\overline{T} $ deformation of quantum field theory as
			random geometry}},  \href{http://dx.doi.org/10.1007/JHEP10(2018)186}{{\sf
			JHEP} {\sf {10} }{\sf (2018) }{\sf 186}},
	\href{http://arxiv.org/abs/1801.06895}{{\ttfamily arXiv:1801.06895
			[hep-th]}}.
	
	\bibitem{Conti:2018tca}
	R.~Conti, S.~Negro, and R.~Tateo, {\it {The $ \mathrm{T}\overline{\mathrm{T}} $
			perturbation and its geometric interpretation}},
	\href{http://dx.doi.org/10.1007/JHEP02(2019)085}{{\sf JHEP} {\sf {02} }{\sf
			(2019) }{\sf 085}}, \href{http://arxiv.org/abs/1809.09593}{{\ttfamily
			arXiv:1809.09593 [hep-th]}}.
	
	\bibitem{Tolley:2019nmm}
	A.~J. Tolley, {\it {$T \bar T$ Deformations, Massive Gravity and Non-Critical
			Strings}},  \href{http://arxiv.org/abs/1911.06142}{{\ttfamily
			arXiv:1911.06142 [hep-th]}}.
	
	\bibitem{Cardy:2019qao}
	J.~Cardy, {\it {$T\bar T$ deformation of correlation functions}},
	\href{http://dx.doi.org/10.1007/JHEP12(2019)160}{{\sf JHEP} {\sf {19} }{\sf
			(2020) }{\sf 160}}, \href{http://arxiv.org/abs/1907.03394}{{\ttfamily
			arXiv:1907.03394 [hep-th]}}.
	
	\bibitem{McGough:2016lol}
	L.~McGough, M.~Mezei, and H.~Verlinde, {\it {Moving the CFT into the bulk with
			$ T\overline{T} $}},  \href{http://dx.doi.org/10.1007/JHEP04(2018)010}{{\sf
			JHEP} {\sf {04} }{\sf (2018) }{\sf 010}},
	\href{http://arxiv.org/abs/1611.03470}{{\ttfamily arXiv:1611.03470 [hep-th]}}.
	%
	
	\bibitem{Kraus:2018xrn}
	P.~Kraus, J.~Liu, and D.~Marolf, {\it {Cutoff AdS$_{3}$ versus the $
			T\overline{T} $ deformation}},
	\href{http://dx.doi.org/10.1007/JHEP07(2018)027}{{\sf JHEP} {\sf {07} }{\sf
			(2018) }{\sf 027}},
	\href{http://arxiv.org/abs/1801.02714}{{\ttfamily arXiv:1801.02714 [hep-th]}}.
	%
	
	\bibitem{Taylor:2018xcy}
	M.~Taylor, {\it {TT deformations in general dimensions}},
	\href{http://arxiv.org/abs/1805.10287}{{\ttfamily arXiv:1805.10287 [hep-th]}}.
	%
	
	\bibitem{Hartman:2018tkw}
	T.~Hartman, J.~Kruthoff, E.~Shaghoulian, and A.~Tajdini, {\it {Holography at
			finite cutoff with a $T^2$ deformation}},
	\href{http://dx.doi.org/10.1007/JHEP03(2019)004}{{\sf JHEP} {\sf {03} }{\sf
			(2019) }{\sf 004}}, \href{http://arxiv.org/abs/1807.11401}{{\ttfamily
			arXiv:1807.11401 [hep-th]}}.
	
	\bibitem{Belin:2020oib}
	A.~Belin, A.~Lewkowycz, and G.~Sarosi, {\it {Gravitational path integral from
			the $T^2$ deformation}},
	\href{http://dx.doi.org/10.1007/JHEP09(2020)156}{{\sf JHEP} {\sf {09} }{\sf
			(2020) }{\sf 156}}, \href{http://arxiv.org/abs/2006.01835}{{\ttfamily
			arXiv:2006.01835 [hep-th]}}.
	
	\bibitem{Kruthoff:2020hsi}
	J.~Kruthoff and O.~Parrikar, {\it {On the flow of states under
			$T\overline{T}$}},  \href{http://arxiv.org/abs/2006.03054}{{\ttfamily
			arXiv:2006.03054 [hep-th]}}.
	
	\bibitem{Li:2020pwa}
	Y.~Li and Y.~Zhou, {\it {Cutoff $\rm AdS_3$ versus $\rm T\bar{T}$ $\rm CFT_2$
			in the large central charge sector: correlators of energy-momentum tensor}},
	\href{http://arxiv.org/abs/2005.01693}{{\ttfamily arXiv:2005.01693
			[hep-th]}}.
	
	\bibitem{Giveon:2017myj}
	A.~Giveon, N.~Itzhaki, and D.~Kutasov, {\it {A solvable irrelevant deformation
			of AdS$_{3}$/CFT$_{2}$}},
	\href{http://dx.doi.org/10.1007/JHEP12(2017)155}{{\sf JHEP} {\sf {12} }{\sf
			(2017) }{\sf 155}},
	\href{http://arxiv.org/abs/1707.05800}{{\ttfamily arXiv:1707.05800 [hep-th]}}.
	%
	
	\bibitem{Guica:2019nzm}
	M.~Guica and R.~Monten, {\it {$T\bar T$ and the mirage of a bulk cutoff}},
	\href{http://arxiv.org/abs/1906.11251}{{\ttfamily arXiv:1906.11251
			[hep-th]}}.
	
	\bibitem{Hirano:2020nwq}
	S.~Hirano and M.~Shigemori, {\it {Random Boundary Geometry and Gravity Dual of
			$T\bar{T}$ Deformation}},  \href{http://arxiv.org/abs/2003.06300}{{\ttfamily
			arXiv:2003.06300 [hep-th]}}.
	
	\bibitem{Apolo:2019zai}
	L.~Apolo, S.~Detournay, and W.~Song, {\it {TsT, $T\bar{T}$ and black strings}},
	\href{http://dx.doi.org/10.1007/JHEP06(2020)109}{{\sf JHEP} {\sf {06} }{\sf
			(2020) }{\sf 109}}, \href{http://arxiv.org/abs/1911.12359}{{\ttfamily
			arXiv:1911.12359 [hep-th]}}.
	
	\bibitem{Zamolodchikov:2004ce}
	A.~B. Zamolodchikov, {\it {Expectation value of composite field T anti-T in
			two-dimensional quantum field theory}},
	\href{http://arxiv.org/abs/hep-th/0401146}{{\ttfamily arXiv:hep-th/0401146
			[hep-th]}}.
	%
	
	\bibitem{Bonelli:2018kik}
	G.~Bonelli, N.~Doroud, and M.~Zhu, {\it {$T\bar T$-deformations in closed
			form}},
	\href{http://arxiv.org/abs/1804.10967}{{\ttfamily arXiv:1804.10967 [hep-th]}}.
	%
	
	\bibitem{Coleman:2019dvf}
	E.~A. Coleman, J.~Aguilera-Damia, D.~Z. Freedman, and R.~M. Soni, {\it {$
			T\overline{T} $ -deformed actions and (1,1) supersymmetry}},
	\href{http://dx.doi.org/10.1007/JHEP10(2019)080}{{\sf JHEP} {\sf {10} }{\sf
			(2019) }{\sf 080}}, \href{http://arxiv.org/abs/1906.05439}{{\ttfamily
			arXiv:1906.05439 [hep-th]}}.
	
	\bibitem{Caputa:2019pam}
	P.~Caputa, S.~Datta, and V.~Shyam, {\it {Sphere partition functions \& cut-off
			AdS}},  \href{http://dx.doi.org/10.1007/JHEP05(2019)112}{{\sf JHEP} {\sf {05}
		}{\sf (2019) }{\sf 112}}, \href{http://arxiv.org/abs/1902.10893}{{\ttfamily
			arXiv:1902.10893 [hep-th]}}.
	
	\bibitem{Jiang:2019tcq}
	Y.~Jiang, {\it {Expectation value of $\mathrm{T}\overline{\mathrm{T}}$ operator
			in curved spacetimes}},
	\href{http://dx.doi.org/10.1007/JHEP02(2020)094}{{\sf JHEP} {\sf {02} }{\sf
			(2020) }{\sf 094}}, \href{http://arxiv.org/abs/1903.07561}{{\ttfamily
			arXiv:1903.07561 [hep-th]}}.
	
	
	\bibitem{Dubovsky:2018bmo}
	S.~Dubovsky, V.~Gorbenko, and G.~Hern{\'a}ndez-Chifflet, {\it {$T\bar{T}$
			Partition Function from Topological Gravity}},
	\href{http://arxiv.org/abs/1805.07386}{{\ttfamily arXiv:1805.07386 [hep-th]}}.
	%
	
	\bibitem{Freidel:2008sh}
	L.~Freidel, {\it {Reconstructing AdS/CFT}},
	\href{http://arxiv.org/abs/0804.0632}{{\ttfamily arXiv:0804.0632 [hep-th]}}.
	%
	
	\bibitem{Mazenc:2019cfg}
	E.~A. Mazenc, V.~Shyam, and R.~M. Soni, {\it {A $T \bar{T}$ Deformation for
			Curved Spacetimes from 3d Gravity}},
	\href{http://arxiv.org/abs/1912.09179}{{\ttfamily arXiv:1912.09179
			[hep-th]}}.
	
	\bibitem{Henningson:1998gx}
	M.~Henningson and K.~Skenderis, {\it {The Holographic Weyl anomaly}},
	\href{http://dx.doi.org/10.1088/1126-6708/1998/07/023}{{\sf JHEP} {\sf {07}
		}{\sf (1998) }{\sf 023}},
	\href{http://arxiv.org/abs/hep-th/9806087}{{\ttfamily arXiv:hep-th/9806087
			[hep-th]}}.
	%
	
	\bibitem{deHaro:2000vlm}
	S.~de~Haro, S.~N. Solodukhin, and K.~Skenderis, {\it {Holographic
			reconstruction of space-time and renormalization in the AdS / CFT
			correspondence}},  \href{http://dx.doi.org/10.1007/s002200100381}{{\sf
			Commun. Math. Phys.} {\sf {217} }{\sf (2001) }{\sf 595--622}},
	\href{http://arxiv.org/abs/hep-th/0002230}{{\ttfamily arXiv:hep-th/0002230
			[hep-th]}}.
	%
	
	\bibitem{Skenderis:1999nb}
	K.~Skenderis and S.~N. Solodukhin, {\it {Quantum effective action from the AdS
			/ CFT correspondence}},
	\href{http://dx.doi.org/10.1016/S0370-2693(99)01467-7}{{\sf Phys. Lett. B}
		{\sf {472} }{\sf (2000) }{\sf 316--322}},
	\href{http://arxiv.org/abs/hep-th/9910023}{{\ttfamily arXiv:hep-th/9910023}}.
	
	\bibitem{Balasubramanian:1999re}
	V.~Balasubramanian and P.~Kraus, {\it {A Stress tensor for Anti-de Sitter
			gravity}},  \href{http://dx.doi.org/10.1007/s002200050764}{{\sf Commun. Math.
			Phys.} {\sf {208} }{\sf (1999) }{\sf 413--428}},
	\href{http://arxiv.org/abs/hep-th/9902121}{{\ttfamily arXiv:hep-th/9902121
			[hep-th]}}.
	%
	
	\bibitem{Grosse:2007vr}
	H.~Grosse and G.~Lechner, {\it {Wedge-Local Quantum Fields and Noncommutative
			Minkowski Space}},
	\href{http://dx.doi.org/10.1088/1126-6708/2007/11/012}{{\sf JHEP} {\sf {11}
		}{\sf (2007) }{\sf 012}}, \href{http://arxiv.org/abs/0706.3992}{{\ttfamily
			arXiv:0706.3992 [hep-th]}}.
	
	\bibitem{Dubovsky:2012wk}
	S.~Dubovsky, R.~Flauger, and V.~Gorbenko, {\it {Solving the Simplest Theory of
			Quantum Gravity}},  \href{http://dx.doi.org/10.1007/JHEP09(2012)133}{{\sf
			JHEP} {\sf {09} }{\sf (2012) }{\sf 133}},
	\href{http://arxiv.org/abs/1205.6805}{{\ttfamily arXiv:1205.6805 [hep-th]}}.
	
\end{thebibliography}\endgroup
\end{document}